\documentclass[prl,reprint,nofootinbib,superscriptaddress,preprintnumbers,showpacs]{revtex4-1}
\usepackage{amsfonts}
\usepackage{amssymb}
\usepackage{amsmath}
\usepackage{graphicx,color}
\usepackage{dcolumn}
\usepackage{epsfig}
\usepackage{bm}
\usepackage{bbm}
\usepackage{ulem}
\usepackage{hyperref}
\usepackage{lineno}

\usepackage{color}

\def\gtap{\ \raise.3ex\hbox{$>$\kern-.75em\lower1ex\hbox{$\sim$}}\ }
\def\ltap{\ \raise.3ex\hbox{$<$\kern-.75em\lower1ex\hbox{$\sim$}}\ }
\begin{document}

\title{
$Z_c(4430)$ and $Z_c(4200)$ as triangle singularities
}
\author{S. X. Nakamura}
\email{satoshi@ustc.edu.cn}
\affiliation{
University of Science and Technology of China, Hefei 230026, 
People's Republic of China
}
\affiliation{
State Key Laboratory of Particle Detection and Electronics (IHEP-USTC), Hefei 230036, People's Republic of China}
\affiliation{
Laborat\'orio de F\'isica Te\'orica e Computacional - LFTC, 
Universidade Cruzeiro do Sul / Universidade Cidade de S\~ao Paulo, S\~ao Paulo, SP 01506-000, Brazil
}
\author{K. Tsushima}
\affiliation{
Laborat\'orio de F\'isica Te\'orica e Computacional - LFTC, 
Universidade Cruzeiro do Sul / Universidade Cidade de S\~ao Paulo, S\~ao Paulo, SP 01506-000, Brazil
}

\preprint{LFTC-19-5/43}

\begin{abstract}
 $Z_c(4430)$ discovered by the Belle
 and confirmed by the LHCb
 in $\bar{B}^0\to\psi(2S)K^-\pi^+$
 is generally considered to be 
 a charged charmonium-like state that includes minimally two quarks and
 two antiquarks.
 $Z_c(4200)$ found in $\bar{B}^0\to J/\psi K^-\pi^+$
 by the Belle is also a good candidate of 
 a charged charmonium-like state.
 In this work, we propose a compelling alternative to the tetraquark-based
interpretations of $Z_c(4430)$ and $Z_c(4200)$.
 We demonstrate that kinematical singularities in triangle loop diagrams induce
a resonance-like behavior that can consistently explain the
 properties (spin-parity, mass, width, and Argand plot) of 
 $Z_c(4430)$ and $Z_c(4200)$ from the experimental analyses.
 Applying this idea to $\Lambda_b^0\to J/\psi p\pi^-$,
 we also identify triangle singularities that behave like $Z_c(4200)$,
 but no triangle diagram is available for $Z_c(4430)$.
 This is consistent with the LHCb's finding that 
 their description of the $\Lambda_b^0\to J/\psi p\pi^-$
 data is significantly improved by including
 a $Z_c(4200)$ contribution while $Z_c(4430)$ seems to hardly contribute.
 Even though the proposed mechanisms have uncertainty in 
 the absolute strengths which are currently difficult to estimate,
otherwise the results are essentially determined by the kinematical
 effects and thus robust.
\end{abstract}

\maketitle

Charged quarkonium-like states, so-called $Z_c$
and $Z_b$~\footnote{We follow Ref.~\cite{pdg}
on the particle notations.},
occupy a special position in the contemporary hadron spectroscopy.
This is because, if they do exist, they clearly consist of
at least four valence (anti)quarks, being different from the conventional
quark-antiquark structure.
The QCD phenomenology would become significantly richer
by establishing their existence.
Among $\sim 10$ of such states that have been claimed to exist
as of 2018, we focus on
$Z_c(4430)$ and $Z_c(4200)$.

$Z_c(4430)$ was discovered by the Belle Collaboration
as a bump in the  
$\psi(2S)\pi^+$ invariant mass distribution
of $\bar B^0\to\psi(2S) K^-\pi^+$~\cite{belle_z4430_2008};
charge conjugate modes are implicitly included throughout.
Many theoretical interpretations of $Z_c(4430)$
have been proposed: diquark-antidiquark~\cite{z4430-tetraquark1,z4430-tetraquark2,z4430-tetraquark3},
hadronic molecule~\cite{z4430-molecule1,z4430-molecule2,z4430-molecule3,z4430-molecule4,z4430-molecule5},
and kinematical threshold cusp~\cite{z4430_cusp1,z4430_cusp2},
as summarized in reviews~\cite{review_hosaka,review_chen,review_lebed,review_raphael}.
The experimental determination of the spin-parity ($J^P=1^+$)
ruled out many of the scenarios~\cite{belle_z4430,lhcb_z4430};
in particular, the threshold cusp has been eliminated. 
After the LHCb Collaboration found a resonance-like
behavior in the $Z_c(4430)$ Argand plot~\cite{lhcb_z4430},
a consensus is that $Z_c(4430)$ is a genuine tetraquark state~\cite{z4430-aps}.
$Z_c(4200)$ is also a good tetraquark candidate~\cite{z4430-tetraquark3,z4200_tetra1}.
It was observed by the Belle in
$\bar B^0\to J/\psi K^-\pi^+$~\cite{belle_z4200}.
The LHCb also found $Z_c(4200)$-like contributions in
$\bar B^0\to J/\psi K^-\pi^+$~\cite{lhcb_z4200}
and $\Lambda_b^0\to J/\psi\, p\,\pi^-$~\cite{lhcb_z4200_Lb}.

\begin{figure}[b]
\begin{center}
\includegraphics[width=.4\textwidth]{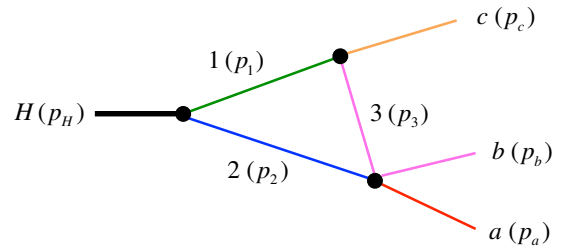}
\end{center}
 \caption{Triangle diagram for $H\to abc$ decay.
 Particle labels and their momenta (in parentheses)
 are defined.
 }
\label{fig:diag2}
\end{figure}

Meanwhile, triangle singularities (TS)~\cite{landau,Aitchison,coleman,schmid,s-matrix}
have been considered to interpret
several resonance(-like) states such as
a hidden charm pentaquark
$P_c(4450)^+$~\cite{TS-Pc,TS-Pc3,TS-Pc2},
and a charged charmonium-like state 
$Z_c(3900)$~\cite{TS-Zc3900-1,TS-Zc3900-2}.
The TS is a kinematical effect that arises in a triangle diagram like Fig.~\ref{fig:diag2}
when a special kinematical condition is reached: three intermediate
particles are,
as in a classical process,
allowed to be on-shell at the same time.
A mathematical detail how the singularity shows up
is well illustrated in Ref.~\cite{TS-Pc2}.
A dispersion theoretical viewpoint is given in Ref.~\cite{TS-DR}.
Although it was claimed in Refs.~\cite{Pakhlov2011,Pakhlov2015} that 
an on-shell triangle loop, which includes
an experimentally unobserved hadron, 
can induce a spectrum bump of $Z_c(4430)$,
the kinematics of the proposed mechanism is in fact classically
forbidden and 
not causing a TS (Coleman-Norton theorem~\cite{coleman};
also see Fig.~4 and related discussion in Ref.~\cite{TS-Pc2}).
The mechanism generates a clockwise Argand plot,
which is opposite to the LHCb data~\cite{lhcb_z4430},
and has already been ruled out~\footnote{We confirmed, within our model described below,
that the triangle diagram of Refs.~\cite{Pakhlov2011,Pakhlov2015}
does not generate a $Z_c(4430)$-like bump.
This is expected from the
Coleman-Norton theorem~\cite{coleman} and 
a general discussion in Ref.~\cite{TS-Pc2}.}.

\begin{figure*}[t]
\begin{center}
\includegraphics[width=1\textwidth]{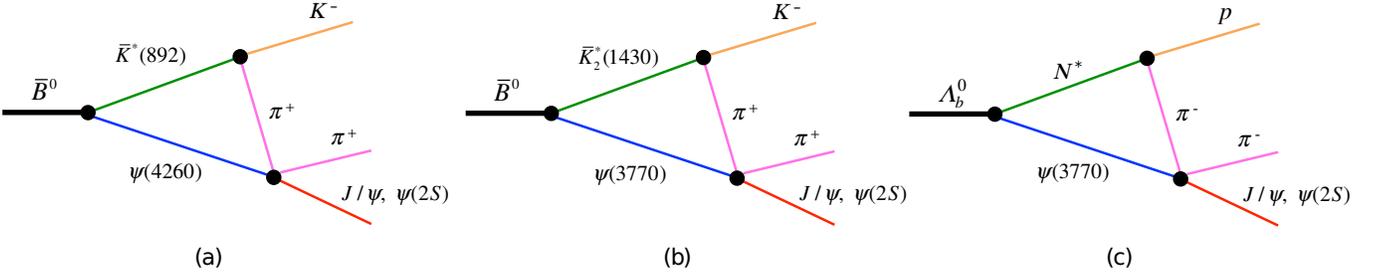}
\end{center}
\caption{Triangle diagrams contributing to 
$\bar B^0\to\psi_f K^-\pi^+$ (a,b) and
$\Lambda_b^0\to\psi_f p\pi^-$ (c);
$\psi_f= J/\psi, \psi(2S)$.
In (c), 
$N^*$ represents an isospin 1/2
nucleon resonances of 1400$-$1800 MeV.
The triangle singularity from
the diagram (a) [(b,c)] generates a
$Z_c(4430)$ [$Z_c(4200)$]-like bump
in the $\psi_f\pi$
invariant mass distribution. 
 }
\label{fig:diag}
\end{figure*}
In this paper,
we give a new insight into $Z_c(4430)$ and $Z_c(4200)$
by showing that these exotic candidates can be consistently
interpreted as TS if the TS have absolute strengths detectable in the
experiments. 
First we point out that triangle diagrams in Fig.~\ref{fig:diag},
formed by experimentally well-established hadrons,
meet the
kinematical condition to cause the TS
(in the zero width limit of unstable particles).
Then we demonstrate that the diagram of 
Fig.~\ref{fig:diag}(a) [Fig.~\ref{fig:diag}(b,c)]
creates a $Z_c(4430)$ [$Z_c(4200)$]-like bump in 
the $\psi_f\pi$ ($\psi_f= J/\psi,\psi(2S)$) invariant mass distribution
of $\bar B^0\to\psi(2S)K^-\pi^+$
[$\bar B^0\to J/\psi K^-\pi^+$ and $\Lambda_b^0\to J/\psi p\pi^-$].
The Breit-Wigner masses and widths fitted to the spectra
turn out to be in very good agreement with those of $Z_c(4430)$ and $Z_c(4200)$.
The $Z_c(4430)$ Argand plot from the LHCb~\cite{lhcb_z4430}
is also well reproduced by the triangle diagram.
Finally, we give a natural explanation for
the absence of $Z_c(4430)$ in $\Lambda_b^0\to J/\psi p\pi^-$
and $e^+e^-$ annihilations in terms of the TS.
This is so far the most successful TS-based interpretation of charged quarkonium-like
states;
$Z_c(3900)$ as TS has been disfavored in
Ref.~\cite{TS-Zc3900-2}~\footnote{
 The TS-based interpretation of $P_c(4450)^+$~\cite{TS-Pc,TS-Pc3,TS-Pc2}
 has been ruled out by recent data~\cite{lhcb-new-penta}.}.

First we show that the triangle diagrams in Fig.~\ref{fig:diag}
hit the TS in the zero width limit of the unstable particles.
A set of equations presented in Sec.~II of Ref.~\cite{TS-Pc2}
is useful for this purpose.
Regarding Fig.~\ref{fig:diag}(a),
we substitute
the PDG averaged particle masses~\cite{pdg}
into the formulas,
and obtain
$p_1=p_2=491$~MeV, $p_3=154$~MeV (the momentum symbols of Fig.~\ref{fig:diag2})
in the $\bar B^0$-at-rest frame, and
$m_{\psi(2S)\pi}=4420$~MeV 
($\psi(2S)\pi$ invariant mass)
at the TS where all particles in the loop have classically allowed
energies and momenta.
Similarly, we obtain $m_{J/\psi\pi}=4187$~MeV at the TS for Fig.~\ref{fig:diag}(b),
and
$m_{J/\psi\pi}=3970$~MeV, 4004~MeV, 4116~MeV for Fig.~\ref{fig:diag}(c)
with $N^*=N(1440)\,1/2^+$, $N(1520)\,3/2^-$, and $N(1680)\,5/2^+$, respectively.
In the realistic case where
the unstable particles have
finite widths,
the triangle diagrams do not exactly hit the TS and
the location of the spectrum peak due to the TS
can be somewhat different from the above $m_{\psi_f\pi}$ values.
Using the same formulas,
we can also confirm that the triangle diagrams of
Refs.~\cite{Pakhlov2011,Pakhlov2015} are, 
in the zero-width limit,
kinematically forbidden at the classical level.

We use a simple and reasonable model to calculate the triangle diagrams
of Fig.~\ref{fig:diag}.
Let us use labeling of particles and their momenta in
Fig.~\ref{fig:diag2} to generally express the triangle amplitudes:
\begin{eqnarray}
 T_{abc,H} &=& \int d\bm{p}_1\,
  { v_{ab;23}(\bm{p}_a,\bm{p}_b;\bm{p}_2,\bm{p}_3)\,
  \Gamma_{3c,1}(\bm{p}_3,\bm{p}_c;\bm{p}_1)
  \over
  E - E_2(\bm{p}_2) - E_3(\bm{p}_3) - E_c(\bm{p}_c)
  }
  \nonumber \\
  &\times&  { 1  \over
  E - E_1(\bm{p}_1) - E_2(\bm{p}_2) }
  \Gamma_{12,H}(\bm{p}_1,\bm{p}_2;\bm{p}_H)
  \ ,
  \label{eq:amp}
\end{eqnarray}
where the summation over spin states of the intermediate particles is
implied.
The quantity $E$ denotes
the total energy in the center-of-mass (CM) frame, 
and
$E_x(\bm{p}_x)=\sqrt{\bm{p}^2_x+m^2_x}$
is the energy of a particle $x$ 
with the mass $m_x$ and momentum $\bm{p}_x$.
An exception is applied to unstable intermediate particles 1
and 2 for which 
$E_j(\bm{p}_j)=m_j + \bm{p}^2_j/2m_j - i\Gamma_j/2\ (j=1,2)$
where $\Gamma_j$ is the width.
It is important to consider
the vector charmonium width in 
Fig.~\ref{fig:diag}(a) where
$\psi(4260)$ and $K^*(892)$ have comparable widths.
We use the mass and width values from Ref.~\cite{pdg}.

Regarding the $23\to ab$ interaction $v_{ab;23}$ in Eq.~(\ref{eq:amp}),
where the particles 2 and $a$ are vector charmoniums while
3 and $b$ are pions,
we use an $s$-wave interaction:
\begin{eqnarray}
 v_{ab;23}(\bm{p}_a,\bm{p}_b;\bm{p}_2,\bm{p}_3)
  = f^{01}_{ab}(p_{ab}) f^{01}_{23}(p_{23})\,
  \bm{\epsilon}^*_a\cdot\bm{\epsilon}_2 \ ,
\label{eq:contact}
\end{eqnarray}
where $\bm{\epsilon}_a$ and $\bm{\epsilon}_2$
are polarization vectors for 
the particles $a$ and 2, respectively.
The form factors
$f^{01}_{ab}(p_{ab})$ and $f^{01}_{23}(p_{23})$
will be defined in Eq.~(\ref{eq:ff});
the momentum of the particle $i$ in the $ij$-CM frame is denoted by
$\bm{p}_{ij}$ and $p_{ij}=|\bm{p}_{ij}|$.
An $s$-wave pair of $\psi_f\pi$
coming out from this interaction has $J^P=1^+$, which is 
consistent with the experimentally determined spin-parity of 
$Z_c(4430)$ and $Z_c(4200)$, and also with the insignificant $d$-wave
contribution in the $Z_c(4430)$-region~\cite{lhcb_z4430}.

\begin{figure*}[t]
\begin{center}
\includegraphics[width=1\textwidth]{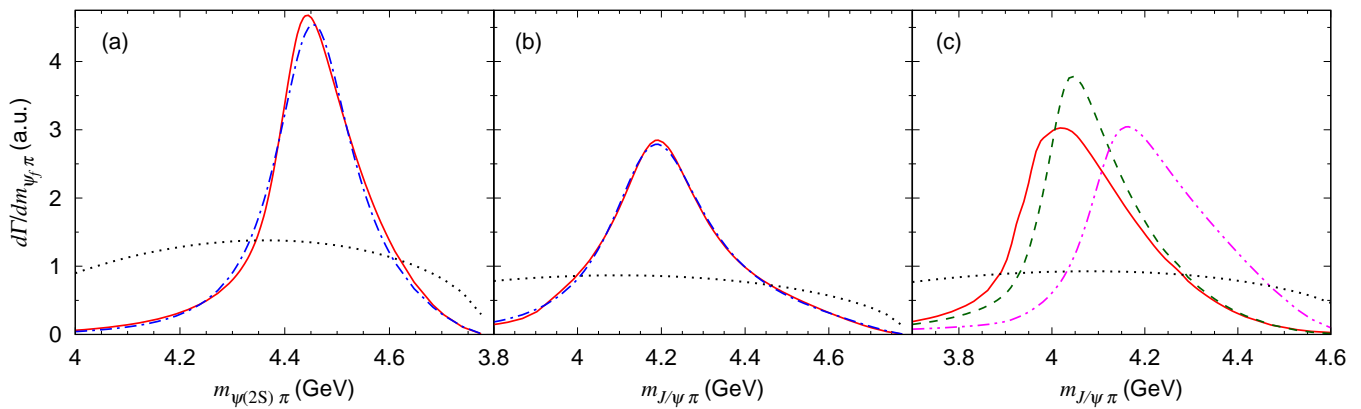}
\end{center}
\caption{
 Distributions of the
 $\psi_f\pi$ ($\psi_f= J/\psi, \psi(2S)$)
invariant mass for 
$\bar B^0\to\psi(2S) K^-\pi^+$ (a),
$\bar B^0\to J/\psi K^-\pi^+$ (b),
and $\Lambda_b^0\to J/\psi p\pi^-$~(c).
The red solid curves in panels (a) and (b) are obtained from
triangle diagrams Figs.~\ref{fig:diag}(a) and \ref{fig:diag}(b), respectively.
The blue dash-dotted curves are from Breit-Wigner amplitudes fitted
 to the red solid curves.
In panel (c),
the red solid, green dashed, and magenta dash-two-dotted curves are obtained from 
Fig.~\ref{fig:diag}(c) with $N^*=N(1440)\,1/2^+$, $N(1520)\,3/2^-$, and 
$N(1680)\,5/2^+$, respectively.
The dotted curves are the phase-space distributions.
 Each curve, except for the blue dash-dotted ones,
 is normalized to give unity when integrated with respect to
$m_{\psi_f\pi}$.
}
\label{fig:spec}
\end{figure*}
The $R\to ij$ decay vertex $\Gamma_{ij,R}$
in Eq.~(\ref{eq:amp})
is explicitly given as
\begin{eqnarray}
  \Gamma_{ij,R}(\bm{p}_i,\bm{p}_j;\bm{p}_R)
   &=& \sum_{LS}f^{LS}_{ij}(p_{ij}) (s_is_i^zs_js_j^z|SS^z)
   \nonumber \\
&\times&   (LM S S^z|s_Rs^z_R)
   Y_{LM} (\hat{p}_{ij}) \ ,
  \label{eq:vertex}
\end{eqnarray}
where $Y_{LM}$ is spherical harmonics.
Clebsch-Gordan coefficients are written as $(abcd|ef)$,
and the spin and its $z$-component of a particle $x$ are
denoted by $s_x$ and $s_x^z$, respectively.
The form factor $f^{LS}_{ij}(p_{ij})$ is parametrized as
\begin{eqnarray}
%
 f^{LS}_{ij}(p) =
g^{LS}_{ij} {p^L\over \sqrt{E_i(p) E_j(p)}}
\left(\frac{\Lambda^2}{\Lambda^2+p^2}\right)^{2+(L/2)}\ ,
\label{eq:ff}
\end{eqnarray}
where we use the same cutoff for all the vertices, and set
 $\Lambda=1$~GeV throughout unless otherwise stated.
For each of the $1\to 3c$ and $23\to ab$ interactions,
there is only one available set of $\{L,S\}$.
We can determine the $g^{LS}_{ij}$ values for the 
$1\to 3c$ interactions using data such as
$\bar K^*(892)/\bar K^*_{2}(1430)\to K^-\pi^+$,
and $N^*\to\pi^- p$ partial decay widths.
One might think
the $23\to ab$ coupling strength can also be determined
using
$2\to ab\bar{3}$ ($\bar{3}$: antiparticle of 3)
partial decay width.
However, the $ab$ invariant mass in the triangle diagram
is significantly larger (by $\gtap$ 500~MeV) than that of the
$2\to ab\bar{3}$ decay process, 
and thus 
the coupling strengths may be very different between the two.
We leave the $23\to ab$ couplings arbitrary.

The $H\to 12$ decay vertices are currently not well understood
because detailed experimental and lattice QCD inputs are lacking. 
There are still some hints to support
the reasonability of considering
the $\bar B^0\to \psi(4260) \bar{K}^*(892)$ vertex in Fig.~\ref{fig:diag}(a):
(i) the Belle found
excess of $B\to \psi(4260) K$ events above the background~\cite{belle_y4260};
(ii) the D0's data can be consistently interpreted that 
some $b$-flavored hadrons 
weakly decay into states including $\psi(4260)$~\cite{d0_y4260}.
Because the details of the $H\to 12$ vertex
would not change the main conclusions,
 we assume simple structures and
use arbitrary strengths.
Among several sets of $\{L,S\}$ available to 
the $\bar{B}^0$ decays,
we set
$g^{LS}_{ij}\ne 0$
only for $S=|s_1-s_2|$ and
the lowest allowed $L$;
$g^{LS}_{ij}=0$ for the other $\{L,S\}$.
Because of using the above $v_{ab;23}$,
the $\bar{B}^0$ decays are necessarily parity-violating.
For the $\Lambda_b^0$ decays, on the other hand,
both parity-conserving and -violating interactions are possible.
We choose the parity-conserving one and
set
$g^{LS}_{ij}\ne 0$
only for 
$S=|s_1-s_2|$ and the lowest allowed $L$;
$g^{LS}_{ij}=0$ otherwise.

We evaluate
the interactions of Eqs.~(\ref{eq:contact}) and (\ref{eq:vertex})
in the CM frame of the two-body subsystem,
and then multiply kinematical factors to account for the Lorentz
transformation to the total three-body CM frame; see
Appendix~C of Ref.~\cite{3pi}.
The procedure of calculating
the Dalitz plot distribution for $H\to abc$ 
using $T_{abc,H}$ of Eq.~(\ref{eq:amp})
is detailed in 
Appendix~B of Ref.~\cite{3pi}.

We first present
the $\psi_f\pi$
invariant mass distributions
for $\bar B^0\to\psi(2S) K^-\pi^+$ and
$\bar B^0\to J/\psi K^-\pi^+$.
The red solid curves in Figs.~\ref{fig:spec}(a) and \ref{fig:spec}(b)
are solely from the triangle
diagrams of Figs.~\ref{fig:diag}(a) and \ref{fig:diag}(b), respectively.
For comparison, we also plot the phase-space distributions by the black
dotted curves.
A clear resonance-like peak appears at 
$m_{\psi(2S)\pi}\sim 4.45$~GeV in Fig.~\ref{fig:spec}(a)
($m_{J/\psi\pi}\sim 4.2$~GeV in Fig.~\ref{fig:spec}(b))
due to the TS.
We also calculated the $m_{J/\psi\pi}$
spectrum for $\bar B^0\to J/\psi K^-\pi^+$
from the triangle diagram of Fig.~\ref{fig:diag}(a),
and obtained a result very similar to 
Fig.~\ref{fig:spec}(a) after the normalization explained in the caption.

\begin{figure}[b]
\begin{center}
\includegraphics[width=.5\textwidth]{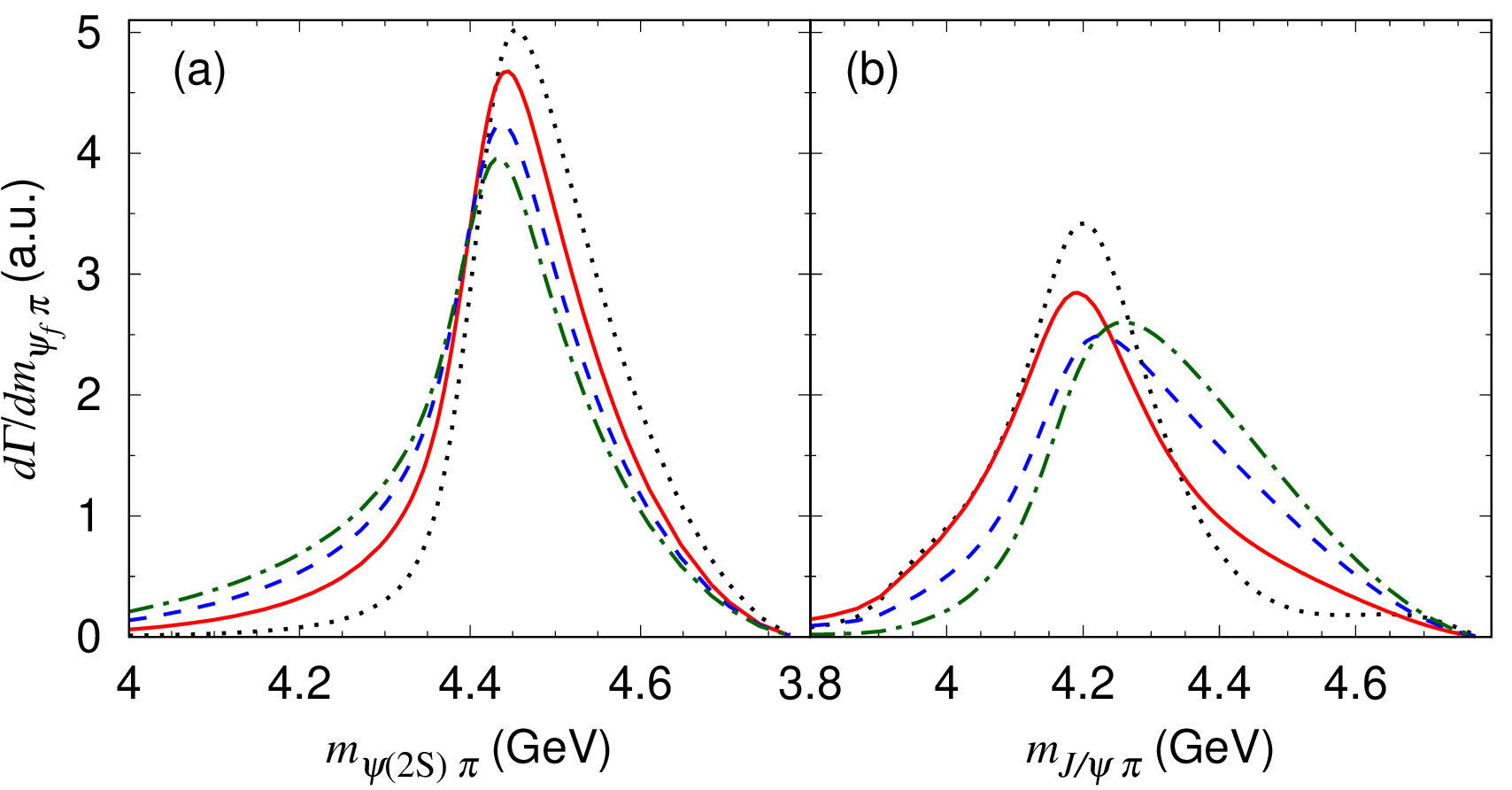}
\end{center}
 \caption{
 Cutoff dependence of the spectra in Fig.~\ref{fig:spec}.
 The panels (a) and (b) correspond to Figs.~\ref{fig:spec}(a) and \ref{fig:spec}(b),
 respectively.
 The red solid curves are the same as those in Fig.~\ref{fig:spec}, and
 are from calculations using the cutoff $\Lambda=1$~GeV.
 The black dotted, blue dashed, and green dash-dotted curves are
 obtained with $\Lambda=$ 0.5, 1.5, and 2~GeV,  respectively.
 All the curves are normalized as in Fig.~\ref{fig:spec}.
 }
\label{fig:cutoff}
\end{figure}
In an ideal situation where experimental inputs are available to
determine all the vertices appearing in the triangle diagrams,
we can make a solid prediction of the spectra to be shown in 
Fig.~\ref{fig:spec}.
This is not the case in reality, and thus
we examine how the above results
depend on the cutoff $\Lambda$
of the form factors in Eq.~(\ref{eq:ff}).
The spectra in Fig.~\ref{fig:cutoff} are obtained by 
changing the cutoff over a reasonable range: $\Lambda=$ 0.5--2~GeV.
The clear peak structures are stable, and the positions and
widths of the bumps do not largely change.
Therefore, we can conclude that
the bump structures in Fig.~\ref{fig:spec}
are essentially determined by the kinematical singularities
and are robust in this reasonable cutoff range.
The stability of the bumps against changing the cutoff can be explained below.
When all particles in the loop have zero widths,
the loop momentum exactly hits the TS at a certain $m_{\psi_f\pi}$,
which blows up the spectrum to infinity
irrespective of the cutoff value.
The finite widths prevent this from happening
and introduce the cutoff dependence to an extent that they push the TS away
from the physical region.

We associate the peaks from the TS
with fake $Z_c$-excitation mechanisms.
We fit the Dalitz plot distributions from the triangle diagrams
of Figs.~\ref{fig:diag}(a) and \ref{fig:diag}(b)
using the mechanism of
$\bar B^0\to Z_c K^-$ followed by $Z_c \to\psi_f \pi^+$.
The $Z_c$ propagation is expressed by the Breit-Wigner form
used in Ref.~\cite{belle_z4430}.
The fitting parameters included in the $Z_c$-excitation mechanisms
are the Breit-Wigner mass, width, and also 
the cutoff in the form factor of Eq.~(\ref{eq:ff}) at the vertices.
In the fit, we consider the kinematical region where the magnitude of
the Dalitz plot distribution is larger than 10\% of the peak height. 
The obtained fits of reasonable quality
are shown by the blue
dash-dotted curves in Figs.~\ref{fig:spec}(a) and \ref{fig:spec}(b).
Because the spectrum shape from the triangle diagrams
is somewhat different from the Breit-Wigner,
their peak positions are slightly different.
\begin{table}[b]
 \caption{\label{tab:BW_param}
Breit-Wigner mass (third row) and width (fourth row)
 for $Z_c(4430)$ and $Z_c(4200)$; the unit is MeV.
 $Z_c(4430)$ [$Z_c(4200)$]
 parameters are fitted to the Dalitz plot distributions for 
$\bar B^0\to\psi(2S) K^-\pi^+$ (a) [$\bar B^0\to J/\psi K^-\pi^+$ (b)]
generated by triangle diagram Fig.~\ref{fig:diag}(a) [\ref{fig:diag}(b)].
The ranges are from the cutoff dependence. 
 The parameters from the experimental analyses
 are also shown; 
the first (second) errors are statistical (systematic). 
}
\begin{ruledtabular}
\renewcommand\arraystretch{1.3}
\begin{tabular}{ccc|cc}
\multicolumn{3}{c|}{$Z_c(4430)$}&\multicolumn{2}{c}{$Z_c(4200)$} \\
  (a) &Belle~\cite{belle_z4430}&LHCb~\cite{lhcb_z4430}
      & (b) &Belle~\cite{belle_z4200} \\\hline
 $4463\pm 13$ &  $4485\pm 22^{+28}_{-11}$ & $4475\pm 7^{+15}_{-25}$& $4233\pm 48$  &  $4196^{+31}_{-29}{}^{+17}_{-13}$\\
 $195\pm 16$  & $200^{+41}_{-46}{}^{+26}_{-35}$ & $172\pm 13^{+37}_{-34}$& $292\pm 56$  & $370\pm 70^{+70}_{-132}$\\
\end{tabular}
\end{ruledtabular}
\end{table}
We fit the Dalitz plot distributions
corresponding to different cutoffs of $\Lambda=0.5-2$~GeV
(Fig.~\ref{fig:cutoff}), and present in Table~\ref{tab:BW_param}
the range of the resulting Breit-Wigner parameters
along with those from experimental data. 
Their agreement is remarkable.

\begin{figure}[t]
\begin{center}
\includegraphics[width=.5\textwidth]{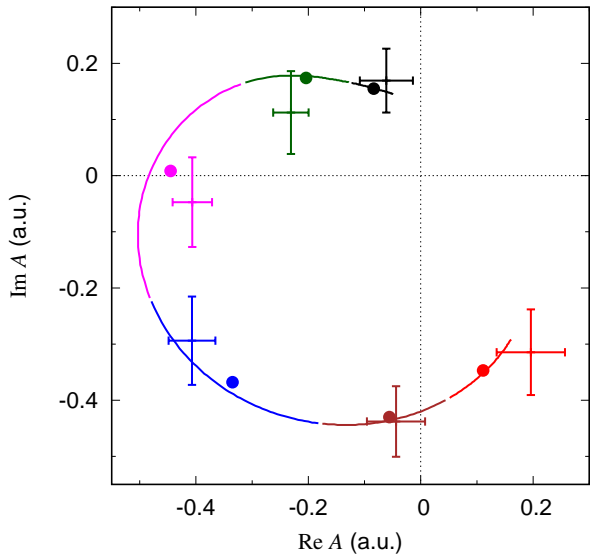}
\end{center}
\caption{
 $Z_c(4430)$ Argand plot.
Six curved segments are
from triangle diagram Fig.~\ref{fig:diag}(a).
Six data points from Ref.~\cite{lhcb_z4430}
are from fitting data in six bins 
equally-separating the range of
 $18~{\rm GeV}^2\le m^2_{\psi(2S)\pi}\le 21.5~{\rm GeV}^2$; 
 $m^2_{\psi(2S)\pi}$ increases counterclockwise.
 A curved segment and a data point of the same color 
 belong to the same bin.
 A solid circle is an average of
 the curved segment of the same color.
 See Eq.~(\ref{eq:average}) for averaging.
 }
\label{fig:argand}
\end{figure}

Next we confront the triangle amplitude
with the $Z_c(4430)$ Argand plot from the LHCb~\cite{lhcb_z4430}.
Because $Z_c$ and $K^-$ are relatively in $p$-wave,
 the angle-independent part of the amplitude ($A$) to be
compared with the Argand plot is
\begin{eqnarray}
A(m^2_{ab}) = c_{\rm \,bg} + c_{\rm\,norm} \int d\Omega_{p_c} Y^*_{1,-s^z_{Z_c}}(-\hat{p}_c) M_{abc,H} \ ,
  \label{eq:argad}
\end{eqnarray}
where $s^z_{Z_c}$ is the $z$-component of the $Z_c$ spin
and $m_{ab}$ the $ab$ invariant mass.
The invariant amplitude $M_{abc,H}$ is related to 
$T_{abc,H}$ of Eq.~(\ref{eq:amp}) through Eq.~(B3) of Ref.~\cite{3pi}.
Complex constants $c_{\rm\,norm}$ and $c_{\rm \,bg}$
are adjusted to fit the empirical Argand plot;
$c_{\rm \,bg}$ represents a background.
In the LHCb analysis,
a complex value representing the $Z_c(4430)$ amplitude is fitted to
dataset in a $m^2_{\psi(2S)\pi}$ bin with a bin size $\Delta$.
To take account of the bin size, 
we simply average our amplitude
without pursuing a theoretical rigor:
\begin{eqnarray}
 \bar{A}(m^2_{ab}(i)) = {1\over\Delta}
\int^{m^2_{ab}(i)+\Delta/2}_{m^2_{ab}(i)-\Delta/2} A(m^2_{ab})\, dm^2_{ab} \ ,
  \label{eq:average}
\end{eqnarray}
where $m^2_{ab}(i)$ is the central value of an $i$-th bin.
As shown in Fig.~\ref{fig:argand},
the empirical $Z_c(4430)$ Argand plot is
fitted well with $\bar{A}(m^2_{ab}(i))$ 
from the triangle diagram of Fig.~\ref{fig:diag}(a);
$c_{\rm \,bg}=0.12+0.03i$ in Eq.~(\ref{eq:argad}).
This demonstrates that the counterclockwise behavior found in
Ref.~\cite{lhcb_z4430} does not necessarily indicate
the existence of a resonance state.
Similar statements have also been made for threshold cusps~\cite{z4430_cusp2,TS-Pc}.
We also confirmed a counterclockwise behavior of
the Argand plot from the triangle diagram of Fig.~\ref{fig:diag}(b), 
as the Belle~\cite{belle_z4200} found 
the $Z_c(4200)$ amplitude to behave so.

A puzzle about $Z_c(4430)$ is its large branching to
$\psi(2S)\pi$ compared with $J/\psi\pi$:
$R^{\rm exp}_{Z_c(4430)}\equiv {\cal B}[Z^+_c(4430)\to\psi(2S)\pi^+]/{\cal B}[Z^+_c(4430)\to J/\psi\pi^+]\sim 11$~\cite{belle_z4430,belle_z4200}.
This can be qualitatively understood
if $Z_c(4430)$ is due to the TS, and
the coupling strength ratio ($c^R_{\psi\pi}$) of 
$\psi(4260)\pi^+\to\psi(2S)\pi^+$ to
$\psi(4260)\pi^+\to J/\psi\pi^+$
interactions of Eq.~(\ref{eq:contact}) is fixed by
$R^{\rm exp}_{\psi(4260)}\equiv {\cal
B}[\psi(4260)\to\psi(2S)\pi^+\pi^-]/{\cal B}[\psi(4260)\to
J/\psi\,\pi^+\pi^-]= (0.11\pm 0.03\pm 0.03)-(0.55\pm 0.18\pm 0.19)$
from four different solutions of Ref.~\cite{Y4260-ratio}.
Because of the large difference in the phase-space available to the final states,
$R^{\rm model}_{\psi(4260)}=0.29\times |c^R_{\psi\pi}|^2$ is
obtained by using Eq.~(\ref{eq:contact}).
In addition, the larger phase-space allows
resonance(-like) $f_0(980)$~\cite{Y4260-f0}
and $Z_c(3900)$~\cite{Y4260-Z3900} to contribute to 
${\cal B}[\psi(4260)\to J/\psi\,\pi^+\pi^-]$ 
by $\sim 40\%$, and thus $R^{\rm model}_{\psi(4260)}\sim 0.17\times |c^R_{\psi\pi}|^2$.
Therefore, the model reproduces 
$R^{\rm exp}_{\psi(4260)}\sim 0.54$ with $|c^R_{\psi\pi}|\sim 1.8$,
and the puzzling $R^{\rm exp}_{Z_c(4430)}\sim 11$ is also reproduced with
the same $|c^R_{\psi\pi}|$.
It is however noted that this discussion is based on the assumption that
$c^R_{\psi\pi}$ is the same for
the $\psi(4260)\pi^+$ scattering at the TS and the $\psi(4260)$ decays.
As discussed earlier, these two processes
are significantly different in the energy,
and thus $c^R_{\psi\pi}$ is not necessarily the same.

Now we discuss 
the $J/\psi\pi$ invariant mass distribution for
$\Lambda_b^0\to J/\psi p\pi^-$
induced by the triangle diagram of Fig.~\ref{fig:diag}(c).
In the $Z_c(4200)$-region, the TS
is expected to create a spectrum bump.
Interestingly, 
several isospin 1/2 nucleon resonances ($N^*$)
of 1400$-$1800 MeV
can contribute to the singularities
and, depending on the mass and width of $N^*$,
the position and width of the bump can vary.
In Fig.~\ref{fig:spec}(c), 
we show results obtained with some representative four-star resonances:
$N^*=N(1440)\,1/2^+$, $N(1520)\,3/2^-$, and $N(1680)\,5/2^+$.
As expected, the triangle diagrams including different $N^*$
generate different spectrum bumps in the $Z_c(4200)$-region.
In reality, these bumps
may coherently interfere with each other
to create a single broad bump.
Also, other charmoniums of 3650-3900~MeV
with coupling to $J/\psi\pi\pi$,
such as $\psi(2S)$ and $\chi_{c1}(3872)$,
could replace $\psi(3770)$ in Fig.~\ref{fig:diag}(c)
to generate TS bumps in the $Z_c(4200)$-region.
The LHCb analysis~\cite{lhcb_z4200_Lb} found that 
the $\Lambda_b^0\to J/\psi p\pi^-$ decay data is significantly better
described by including the $Z_c(4200)$ amplitude.
Because of limited statistics,
the mass and width of 
$Z_c(4200)$ were assumed to be the same as those in
$\bar B^0\to J/\psi K^-\pi^+$~\cite{belle_z4200}.
Therefore,
the spectrum bumps shown in 
Fig.~\ref{fig:spec}(c), some of which extend to the lower end of
the $Z_c(4200)$-region, are still consistent with 
the LHCb's finding.

Another important finding in the LHCb analysis~\cite{lhcb_z4200_Lb} is that
$Z_c(4430)$ seems to hardly contribute to $\Lambda_b^0\to J/\psi p\pi^-$.
If $Z_c(4430)$ found in $\bar B^0\to\psi(2S) K^-\pi^+$
is due to the TS,
a natural explanation follows:
within experimentally observed hadrons,
no combination of a charmonium and a nucleon resonance is available
to form a triangle diagram like Fig.~\ref{fig:diag}(c)
that causes TS at the $Z_c(4430)$ position.
This idea can be further generalized.
At present, a puzzling situation about $Z_c$ is that
those observed in $e^+e^-$ annihilations
and in $B$ decays are mutually exclusive.
If the $Z_c$ states are due to TSs, the answer is simple:
a TS in a $B$ decay does not exist or is highly suppressed in $e^+e^-$ annihilations, and vice
 versa. 
Therefore, a key to establishing a genuine tetraquark state is to
identify it in different processes including different initial states.
However, there are still cases where,
as we have seen in Figs.~\ref{fig:spec}(b) and \ref{fig:spec}(c),
different TS could induce similar resonance-like behaviors.

In summary, we demonstrated that
$Z_c(4430)$ and $Z_c(4200)$, which are often regarded as genuine
tetraquark states,  
can be consistently interpreted as kinematical singularities from
the triangle diagrams we identified.
The Breit-Wigner parameters fitted to the TS-induced
spectrum bumps of $\bar B^0\to\psi_f K^-\pi^+$ 
are in very good agreement with
those of $Z_c(4430)$ and $Z_c(4200)$ from the Belle and LHCb analyses. 
The $Z_c(4430)$ Argand plot from the LHCb is also well
reproduced.
We also explained in terms of TS why $Z_c(4200)$-like contribution was
observed in $\Lambda_b^0\to J/\psi p\pi^-$ but $Z_c(4430)$ was not.
These results are robust because they are essentially determined by the
kinematical effect, and not sensitive to uncertainty of dynamical details.

\begin{acknowledgments}
The authors thank
A.A. Alves Jr, M.~Charles, T.~Skwarnicki, and G.~Wilkinson
 for detailed information on the $Z_c(4430)$
 Argand plot in Ref.~\cite{lhcb_z4430}.
This work is in part supported by 
National Natural Science Foundation of China (NSFC) under contracts 11625523,
and Funda\c{c}\~ao de Amparo \`a Pesquisa do Estado de S\~ao Paulo (FAPESP),
 Process No.~2016/15618-8, No. 2017/05660-0, 
and the Conselho Nacional de Desenvolvimento Cient\'ifico e Tecnol\'ogico - CNPq, 
Process No. 400826/2014-3, No. 308088/2015-8,
No. 313063/2018-4, No. 426150/2018-0,
and Instituto Nacional de Ci\^encia e Tecnologia - Nuclear Physics 
and Applications (INCT-FNA), Brazil, Process No. 464898/2014-5.
\end{acknowledgments}



\end{document}